\documentclass[%
reprint,
superscriptaddress,
preprintnumbers,
nofootinbib,
amsmath,amssymb,
aps,
prd,
floatfix,
longbibliography
]{revtex4-2}

\usepackage{enumitem}
\usepackage{hyperref}
\usepackage{graphicx}
\usepackage{dcolumn}
\usepackage{bm}
\usepackage{physics}

\usepackage{epsfig}
\usepackage{amsmath}
\input{epsf}
\usepackage{psfrag}
\usepackage{xcolor}
\usepackage{pdfpages}
\usepackage{microtype}
\usepackage{booktabs}
\usepackage[normalem]{ulem}
\usepackage{units}
\usepackage{listings} %
\makeatletter
\AtBeginDocument{\let\LS@rot\@undefined}
\makeatother
\newcommand{\bea}{\begin{eqnarray}}
\newcommand{\eea}{\end{eqnarray}}

\graphicspath{{./}{Figs/}}

\newcommand{\eq}[1]{Eq.~\ref{eq:#1}}
\newcommand{\fig}[1]{Fig.~\ref{fig:#1}}
\newcommand{\tab}[1]{Tab.~\ref{tab:#1}}

\newcommand{\sect}[1]{Sec.~\ref{sec:#1}}

\preprint{xxxx}
\begin{document}
\title{Parton spin correlations and $\mathcal{CP}$ properties in Higgs boson decay at future lepton colliders}
\author{Yi-Lin Wang}
\email{wangyilin@mail.ustc.edu.cn}
\affiliation{Interdisciplinary Center for Theoretical Study,
University of Science and Technology of China, Hefei, Anhui 230026, China}
\affiliation{Peng Huanwu Center for Fundamental Theory, Hefei, Anhui 230026, China}
\affiliation{
Institute of High Energy Physics, Chinese Academy of Sciences, Beijing 100049, China}
\author{Jun Gao}
\email{jung49@sjtu.edu.cn}
\affiliation{State Key Laboratory of Dark Matter Physics, Shanghai Key Laboratory for Particle Physics and Cosmology,
Key Laboratory for Particle Astrophysics and Cosmology (MOE),\\
School of Physics and Astronomy, Shanghai Jiao Tong University, Shanghai 200240, China}
\author{Ying-Ying Li}
\email{liyingying@ihep.ac.cn}
\affiliation{
Institute of High Energy Physics, Chinese Academy of Sciences, Beijing 100049, China}
\author{Hua Xing Zhu}
\email{zhuhx@pku.edu.cn}
\affiliation{School of Physics, Peking University, Beijing 100871, China}
\affiliation{Center for High Energy Physics, Peking University, Beijing 100871, China}
\date{\today}

\begin{abstract}
We present a phenomenological study of partonic spin correlations and $\mathcal{CP}$ properties in $H\to gg$ decay channel at future lepton colliders. We investigate two classes of observables: Lund observable defined based on subjets and four-point energy-energy correlator (E4C) between particles inside two jets. Our results show that the E4C with energy weighted to the power of \(n=4\) achieves the strongest sensitivity to the spin correlations of gluons from Higgs boson decay. Under the assumption of ideal identification of different gluon splitting modes, we estimate that future lepton colliders operating at \(\sqrt{s}=240~\mathrm{GeV}\) with an integrated luminosities of \(5.6~\mathrm{ab}^{-1}\) can successfully probe gluon spin correlations, while \(20~\mathrm{ab}^{-1}\) of data can probe the $\mathcal{CP}$-mixing angle in the \(Hgg\) coupling to \(\lesssim 0.03\pi\) using E4C. We outline strategies for extending this framework to realistic detector-level analyses, which can provide a new pathway for the precision test of Standard Model and searches for new physics.
\end{abstract}

\maketitle

\unitlength = 1mm

\section{Introduction}
Spin correlation are widely investigated at colliders for precision measurements of the Standard Model (SM) and searches for physics beyond the SM (BSM), as well as to test the quantum nature (entanglement and test Bell inequalities) of microscopic interactions at high-energy colliders \cite{Barr:2024djo, Fang:2024ple}. These studies usually consider measuring spin correlations through the decay products of heavy particles. For example, as the lifetime of top quark is much shorter than the time scale of hadronization, the spin correlation of top quark pairs can be measured through the kinematics of the decay products, which has been used to determine the $\mathcal{CP}$ properties of top quark Yukawa, searches for top squarks and heavy Higgs bosons \cite{ATLAS:2014abv, Kobakhidze:2014gqa, Bernreuther:2015yna, Arco:2025ydq, Maltoni:2024tul}. Recently, great progress has been made in probing the effects of quantum entanglement using the spin correlations between top quark pair at the Large Hadron
Collider (LHC) \cite{ATLAS:2023fsd, CMS:2024pts}.

For partons, including gluons and light quarks, which evolve into QCD jets through  showering and hadronization on a timescale of \(\unit[10^{-24}]{s}\), extracting their spin correlations requires a detailed understanding of how such correlations are preserved or modified during jet formation. This is a longstanding challenge in collider physics that demands sustained theoretical and computational efforts. To this end, parton showers including spin correlations up to next-to-leading logarithmic (NLL) accuracy are recently being developed \cite{Dasgupta:2020fwr,Forshaw:2020wrq,Nagy:2020rmk,Karlberg:2021kwr,Herren:2022jej,Preuss:2024vyu,Assi:2023rbu}. Complementarily, quantum approaches to simulating QCD branching in real time have been proposed \cite{Bauer:2019qxa,Bepari:2021kwv, Bauer:2023ujy}, aiming to perform coherent summations in parton showers beyond the perturbative approximation.

Simultaneously, it is crucial to identify and optimize the level of quantum correlations that current and future experiments can realistically probe and utilize these information to probe BSM physics. In this work, we initiate a study of parton spin correlations in hadronic Higgs decays, focusing primarily on the decay channel \(H\to gg\). To probe these correlations, we investigate two spin-sensitive observables: a Lund-plane--based variable defined at the jet level~\cite{Karlberg:2021kwr}, and the four-point energy-energy correlator (E4C) constructed from particles inside jets. We note that energy correlators have been used to study QCD spin physics, see e.g. \cite{Chen:2020adz,Li:2023gkh,Kang:2023big,Mantysaari:2025mht,Huang:2025ljp,Song:2025bdj,Gao:2025cwy,Cao:2025icu,Gao:2025evv}. Our work represents the first application of E4C to probe parton spin correlations in Higgs boson decay and Beyond Standard Model physics. For a review of other applications of energy correlators, we refer to \cite{Moult:2025nhu}.
We consider the \(ZH \to \nu\bar{\nu}H\) production channel at future lepton colliders \cite{FCC-ee, CEPCStudyGroup:2018ghi}, which provides a clean environment that avoids the need for jet reclustering when constructing the E4C observable. 

This work is organized as follows. We first discuss, in Sec. \ref{sec:theory}, the theoretical framework of \(H\to gg\) with a $\mathcal{CP}$-mixing phase incorporated, and introduce the spin-sensitive observables. In \sect{shower}, we present the strategies employed to preserve spin correlations between partons in the parton shower at leading order, and compare the correlation strength to that predicted by matrix element calculations. Building on this, in \sect{pheno} we present the sensitivity to probe spin correlations and test the $\mathcal{CP}$ properties of the \(Hgg\) coupling at future lepton colliders. Finally, in \sect{conclusion} we summarize our findings and outline strategies for future improvements.

\section{Theoretical Framework}
\label{sec:theory}
We consider the spin correlations of gluons in $H \to gg$ decay. As the spin correlations can be sensitive to the $\mathcal{CP}$ properties of $Hgg$ coupling, we parametrize the Higgs effective interaction by introducing a $\mathcal{CP}$-mixing phase $\Delta$:
\begin{align}
    \mathcal{L}_{eff} & =  \frac{1}{4}\frac{\alpha_s}{3 \pi} \frac{h}{v}(\cos{\Delta} \ G^a_{\mu \nu}G^{a, \mu \nu} + \sin{\Delta} \ G^a_{\mu \nu}\tilde{G}^{a, \mu \nu})H.
    \label{eq:Lhgg}
\end{align}
Here, $G^a_{\mu\nu}$ is the reduced gluon field strength tensor, which is invariant under $\mathcal{CP}$ transformations, and $\tilde{G}^a_{\mu\nu}$ is its dual, which is $\mathcal{CP}$-odd. 
The matrix element of $H \rightarrow gg$ is given by:
\begin{align}
    \mathcal{M}_{H\rightarrow gg}^{h_1h_2} &=-\frac{1}{4}\frac{\alpha_s}{3\pi v}\bigg(\epsilon_{\mu\nu\alpha\beta}k_1^{\mu}k_2^{\nu}\epsilon_{h_1}^{*\alpha}\epsilon_{h_2}^{*\beta}\sin\Delta\\
    &-\left[(k_1\cdot k_2)(\epsilon^*_{h_1}\cdot\epsilon^*_{h_2})-(k_1\cdot\epsilon^*_{h_2})(k_2\cdot\epsilon^*_{h_1})\right]\cos\Delta\bigg)\notag.
    \label{eq:amp_hgg}
\end{align}
for gluons with momenta $p_1$ and $p_2$, helicities $h_1$ and $h_2$, and corresponding polarization vectors $\epsilon_{h_1}$ and $\epsilon_{h_2}$, respectively. 
In the Higgs boson rest frame, \eq{amp_hgg} takes the form:
\begin{align}
    \mathcal{M}_{++}  = -\frac{1}{4}\frac{\alpha_s}{3\pi v}\frac{m_H^2}{2}e^{i\Delta}, \,\,
    \mathcal{M}_{--}  = -\frac{1}{4}\frac{\alpha_s}{3\pi v}\frac{m_H^2}{2}e^{-i\Delta} . 
\end{align}
where $\mathcal{M}_{++}$ ($\mathcal{M}_{--}$) represents the matrix element for the two gluons produced with helicity $++(--)$. The amplitude where gluons are of different helicity vanishes due to angular momentum conservation. $\mathcal{CP}$ transformation flips the helicity of gluon, we have $\mathcal{M}_{++} \to \mathcal{M}_{--}$. Thus the amplitude $\mathcal{M}_{++} + \mathcal{M}_{--} \propto \cos\Delta$ is $\mathcal{CP}$ even, while $\mathcal{M}_{++} -\mathcal{M}_{--} \propto \sin\Delta$ is $\mathcal{CP}$ odd.

\subsection{Observables}
Spin correlations of partons have been studied through jet substructures by constructing the azimuthal difference $\phi$ between the planes of splittings in two separate jets, known as Lund observable \cite{Karlberg:2021kwr}:
\begin{align}
    \phi & \equiv \arccos\frac{(\vec{p}_1\times\vec{p}_2)\cdot(\vec{p}_3 \times \vec{p}_4)}{(\abs{\vec{p}_1\times\vec{p}_2})(\abs{\vec{p}_3\times\vec{p}_4})}
    \label{eq:phi}
\end{align}
where $\vec{p}_1$ and $\vec{p}_2$ are the momentum for the two subjets within one jet with $\vec{p}_1>\vec{p}_2$, while $\vec{p}_3$ and $\vec{p}_4$ for the two subjets within the other jet with $\vec{p}_3>\vec{p}_4$.

\begin{figure}
    \centering
    \includegraphics[width=0.95\linewidth]{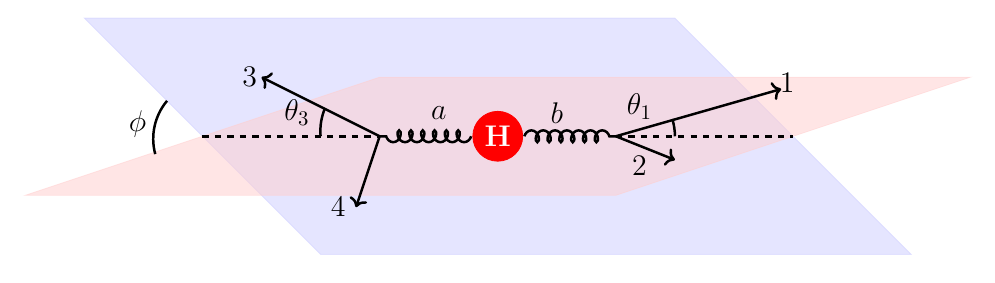}
    \caption{The definition of azimuthal difference $\phi$ in Higgs rest frame for $H\to gg$ at parton level.}
    \label{fig_hgg}
\end{figure}

An azimuthal difference $\phi_{(ij,kl)}$ can also be defined at the level of individual particles within two separated jets, where $(ij)$ and $(kl)$ denote pairs of particles originating from separated jets. Building on this, one can construct an infrared-safe observable sensitive to spin correlations: the four-point energy-energy correlator (E4C), corresponding to the azimuthal difference weighted by the product of the energies of the four particles raised to a power $n \geq 1$~\cite{Chen_2021}:
\begin{align}
    {\rm E4C} (\phi, n) & = \sum _{ij, kl}\int \frac{d \sigma}{\sigma} \frac{E^n_i E^n_j}{Q_1^{2n}} \frac{E^n_k E^n_l}{Q_2^{2n}} \delta \left(\phi- \phi_{(ij, kl)}\right).
    \label{eq:enc}
\end{align}
Here $\{E_i, E_j\}$ correspond to energy of particle $\{i, j\}$ in one jet with jet energy $Q_1$, $\{E_k, E_l\}$ for particle $\{k,l\}$ in another jet with jet energy $Q_2$. The E4C is computed by summing over all such pairs $\{i,j\}$ and $\{k,l\}$ that define the same $\phi$. Note that the calculation of the energy correlator uses the differential cross section defined in the lab frame, whereas the azimuthal angle is measured in the Higgs boson rest frame. 

\section{Parton shower at leading order}
\label{sec:shower}
It is yet to fully understand how the correlation information propagation in showering and jet fragmentation. As a leading-order approximation, we estimate gluon spin correlations in Higgs decay at the parton level by incorporating a one-step gluon splitting. In this approach, the two partons from the splitting are used to approximate subjets for the Lund observable and to approximate the jet-level energy correlator. 
This framework allows for an analytical calculation using spin-dependent splitting function, see e.g. \cite{Catani:1999ss}.

\subsection{Splitting Function}
Consider the splitting $g_1\to a_1 b_1$ and $g_2 \to a_2 b_2$, where $a_i b_i$ can be quark anti-quark pair or a pair of gluons. Eq. (\ref{eq:fac}) gives the factorization for the primary splitting process of the two gluons coming from $H\rightarrow g_1 g_2$.
\begin{align}
    |\mathcal{M}|^2 \simeq &\frac{z_1(1-z_1)}{k^2_{1\perp}} \frac{z_2(1-z_2)}{k^2_{2\perp}}\mathcal{M}^{h_1h_2}_{H\to gg} (\mathcal{M}^{h'_1h'_2}_{H\to gg})^{*} \notag\\
    &D^{h_1h'_1}_{g_1\to a_1b_1}(z_1, k_{1\perp}) D^{h_2h'_2}_{g_2\to a_2b_2}(z_2, k_{2\perp}).
    \label{eq:fac}
\end{align}
Here $\mathcal{M}^{h_1h_2}_{H\to gg}$ is defined in \eq{amp_hgg}. $z_i$ is the energy fraction carried by parton $a_i$ from gluon $g_i$ with helicity $h_i$, and $k_{i\perp}$ is the small transverse momenta.
$D^{h_ih'_i}_{g_i\to a_ib_i}$ is the decay matrix of gluon $g_i$ in the collinear limit:
\begin{align}
    D^{h_ih'_i}_{g_i\to a_ib_i}(z_i,k_{i\perp})&=\epsilon_{h_i,\mu}\epsilon^*_{h'_i,\nu}P^{\mu\nu}_{a_i b_i\leftarrow g_i}(z_i,k_{i\perp}),
\end{align}
and gluon splitting kernel $P^{\mu \nu}_{ab\leftarrow g}$ is given by the Altarelli-Parisi functions \cite{Gribov:1972ri, Dokshitzer:1977sg, ALTARELLI1977298}:
\begin{align}
    P^{\mu \nu}_{q\bar{q}\leftarrow g}(z, k_\perp)  =& P^{\mu \nu}_{\bar{q}q\leftarrow g}(z, k_\perp) \notag\\
    =& {T_F} \left[ -g^{\mu \nu} + 4 z (1 - z) \frac{k_\perp^\mu k_\perp^\nu}{k_\perp^2} \right], \nonumber\\
   P^{\mu \nu}_{gg\leftarrow g}(z, k_\perp)  =&  2C_A \bigg[ -g^{\mu \nu} \left( \frac{z}{1 - z} + \frac{1 - z}{z} \right) \notag\\
   &- 2 z (1 - z) \frac{k_\perp^\mu k_\perp^\nu}{k_\perp^2} \bigg].
    \label{eq:splitting_gluon}
\end{align}

For quark splitting, there is no spin dependence for a collinear splitting, leaving a $\delta_{s s'}$ factor where $s (s')$ denotes the spin of the fermion, as shown below:
\begin{align}
    P^{ss'}_{qg\leftarrow q}(z, k_\perp) & = P^{ss'}_{\bar{q}g\leftarrow \bar{q}}(z, k_\perp) = \delta_{ss'} C_F \frac{1 + z^2}{1 - z}, \nonumber\\
    P^{ss'}_{gq\leftarrow q}(z, k_\perp) & = P^{ss'}_{g\bar{q}\leftarrow q}(z, k_\perp) = \delta_{ss'} C_F  \frac{1 + (1 - z)^2}{z}.
    \label{eq:splitting_quark}
\end{align}

Using \eq{fac}–\eq{splitting_gluon}, we can compute the Lund observables and the E4C for events of the process $H \rightarrow gg$. At leading order, the predictions obtained using the splitting functions should agree in the collinear limit with those from the matrix element calculation performed by \texttt{MadGraph5\_aMC@NLO} (MadGraph), where the tree-level amplitude for $H \to gg \to a_1 b_1 a_2 b_2$ can be evaluated. To carry out this comparison, we simulate the process $e^+e^- \to ZH$ with $H \to gg \to a_1 b_1 a_2 b_2$ and $Z \to \nu\bar{\nu}$ in MadGraph, with the understanding that the intermediate gluons ($gg$) are treated as off-shell.  The transverse momentum $p_{T}$ of final state quarks are required to satisfy $p_{T}\geq 0.1$GeV. Using the truth-level information of the four final-state partons, we reconstruct the Higgs boson kinematics and boost the system to the Higgs rest frame. In the Higgs boson rest frame, to identify the two parton pairs originating from successive gluon splittings, we consider all possible pairings of the four partons and select the configuration in which the openning angle $\Delta R$ between the two partons inside each pair to be smaller than a chosen threshold $\Delta R_{\rm th}$.

\begin{figure}
    \centering
    \includegraphics[width=0.46\textwidth]{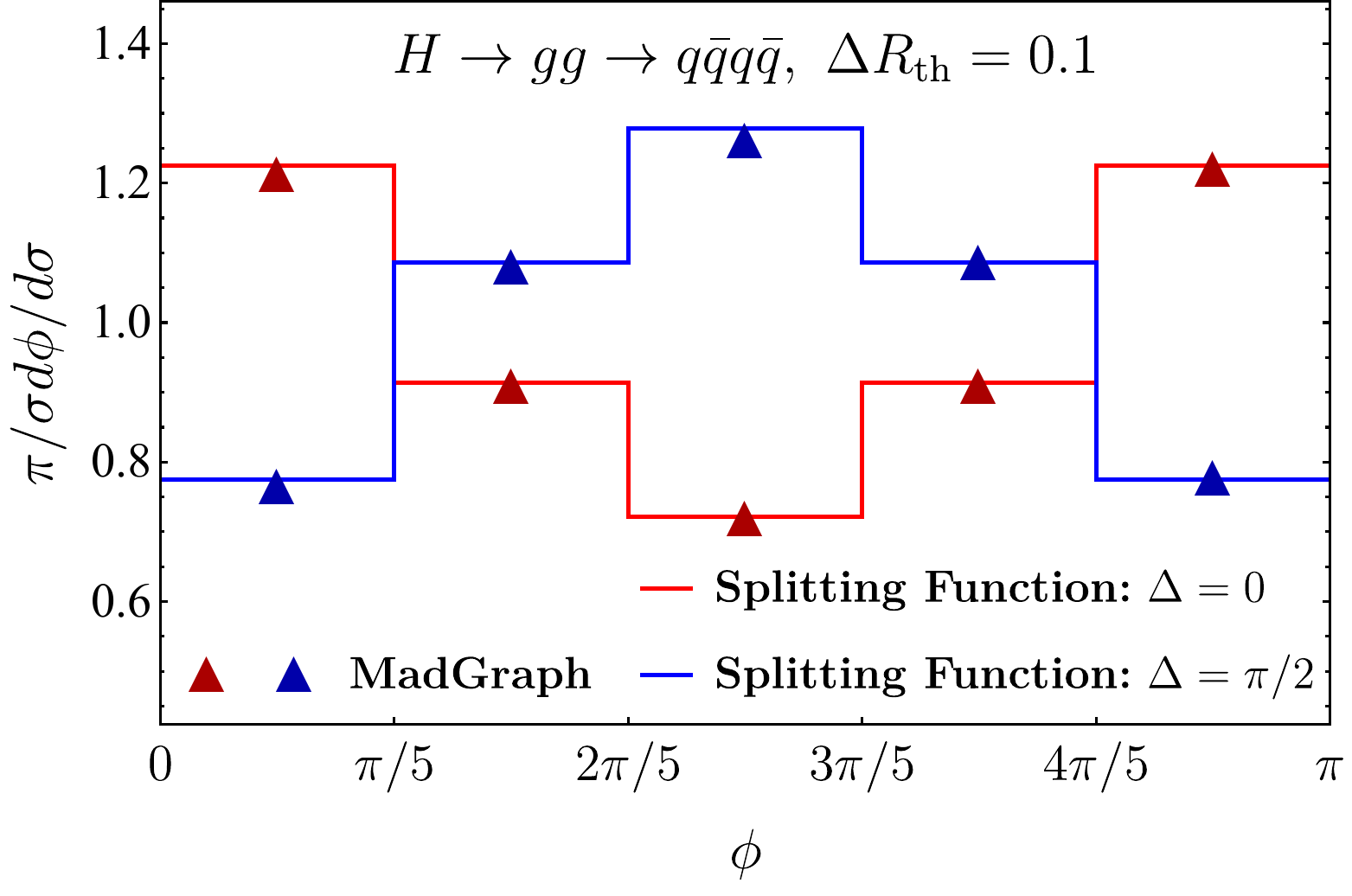}
    \caption{The normalized distribution of the observable $\phi$ for $H\to gg\to q\bar{q}q\bar{q}$ using splitting function (solid line), full matrix element calculations using MadGraph.}
    \label{fig:qqqqn0}
\end{figure}

For the specific case of $H \to gg \to q\bar{q}q\bar{q}$ and the choices of $\Delta R_{\rm th} = 0.1$, we present in \fig{qqqqn0} the normalized distribution $\pi/\sigma d\sigma/d\phi$, which corresponds to the Lund observable as we will discuss later. The distribution obtained from MadGraph is shown as triangle markers. The corresponding distribution computed using the splitting functions is shown as solid lines, where good agreement with that from the matrix element calculation is observed. Furthermore, the distribution doesn't change much for choices of larger $\Delta R_{\rm th}$, as the dominant contribution is from the soft region.

\subsection{E4C \textit{v.s.} Lund observables}
We come to investigate the effectiveness of Lund observables and E4C in revealing the spin correlations of gluon pair from Higgs decay. The Lund observable is defined in the region with $z>z_{\rm cut}$ to ensure infrared safety, where $z$ is the energy fraction carried by the softer parton from the splitting. Using the splitting–function approach, we obtain for $H\!\to gg$ the distribution of the Lund observable $\phi$,
\begin{equation}
    {\rm Obs}(\phi,x)
    \equiv \frac{\pi}{\sigma}\,\frac{d\sigma}{d\phi}
    = 1 + A_{\rm Lund}(x)\cos(2\phi - 2\Delta),
\end{equation}
with the parameter choice $x = z_{\rm cut}$.  
Similarly, the ${\rm E4C}(\phi,x)$ distribution can be written in the
normalized form
\begin{equation}
    {\rm Obs}(\phi,x)
    \equiv \frac{{\rm E4C}(\phi,x)}{\mathcal{N}(x)}
    = 1 + A_{\rm E4C}(x)\cos(2\phi - 2\Delta),
\end{equation}
with $x=n$ and
$\mathcal{N}(x)=\int_{0}^{\pi} d\phi\, {\rm E4C}(\phi,x)$.  
In both cases, the term $\cos(2\phi-2\Delta)$ encodes the expected
spin–correlation structure, while $A_{\text{Lund}/\text{E4C}}(x)$ quantifies the strength of the
correlation for the parameter choice $x$.

We present the correlation strengths $A_{\text{Lund}}(x)$ for two benchmark values of $z_{\rm cut}$ and $A_{\text{E4C}}(x)$ for two
benchmark values of $n$ in \tab{correlation}. Results are shown for the different splitting modes $X \in \{ gggg,\; gg q\bar q,\; q\bar q q\bar q,\;
jjjj \}$,where $q$ denotes the set of light quarks $u,c,d, s$, and $j$ denotes the inclusive mode combining gluons and the light quarks.
\begin{table}[h!]
\centering
\begin{tabular}{|c|c|c|c|c|}
\hline
correlation strength & $q\bar{q}q\bar{q}$  & $gggg$ & $q\bar{q}gg$ & $jjjj$\\
\hline
$A_{\text{Lund}}(z_{\rm cut}=0.1)$ & 0.4204 & 0.0028 & -0.0346 & 0.0003 \\
\hline
$A_{\text{Lund}}(z_{\rm cut}=0.4)$ & 0.9481 & 0.0115 & -0.1044 & 0.0011 \\
\hline
$A_{\text{E4C}}(n=1)$ & 0.4516 & 0.0027 & -0.0349 & 0.0003 \\
\hline
$A_{\text{E4C}}(n=4)$ & 0.6944 & 0.0070 & -0.0695 & 0.0007 \\
\hline
\end{tabular}
\caption{Correlation strength $A_{\text{Lund}}(x)$ for two benchmark values of $z_{\rm cut} = 0.1, 0.4$ and $A_{\text{E4C}}(x)$ for two
benchmark values of $n = 1, 4$.} 
\label{tab:correlation}
\end{table}

We observe that $A_{\text{Lund}}(x)$ 
increases with $z_{\rm cut}$, where the energy of the softest particle becomes larger. This behavior is consistent with the expectation from \eq{splitting_gluon}: the term governing the spin correlation between the two gluons from Higgs decay is proportional to $z(1-z)$, which is maximized at $z = 0.5$ for both the $g \to gg$ and $g \to q\bar{q}$ splittings. Since the E4C is an energy-weighted observable that receives enhanced contributions from regions in which both particles resulting from a splitting are hard, it is natural that $A_{\text{E4C}}(x)$ increases from $n=1$ to $n=4$, as observed \tab{correlation}. This increase is observed to saturate for larger $n$, and we therefore restrict our discussion to $n \le 4$. For the $g \to gg$ splitting mode, this trend is further supported by expectations from the soft theorem \cite{Larkoski:2014bxa}: soft radiation of massless gauge bosons obeys universal soft limits in which spin information is retained by the hard particles. Consequently, the E4C suppresses contributions from events where spin correlations fail to propagate through splittings dominated by soft emissions.
The same reasoning applies to the case of Lund observable: increasing $z_{\rm cut}$ removes an even larger fraction of soft configurations, thereby enhancing the contribution from harder splittings and leading to a larger value of $A_{\text{Lund}}(x)$.

\begin{figure*}
    \centering
    \begin{tabular}{cc}
        \includegraphics[width=0.46\textwidth]{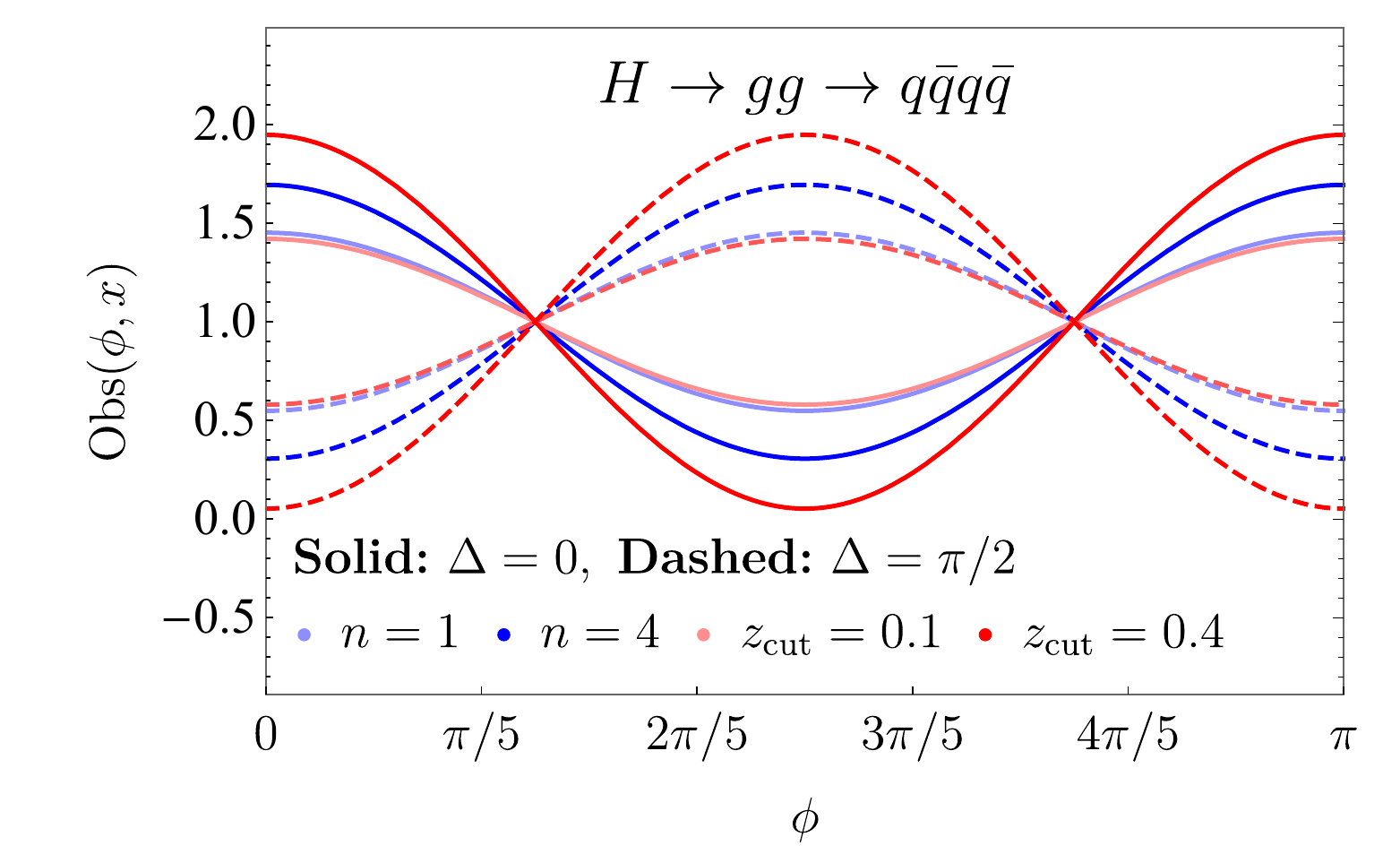} & \includegraphics[width=0.46\textwidth]{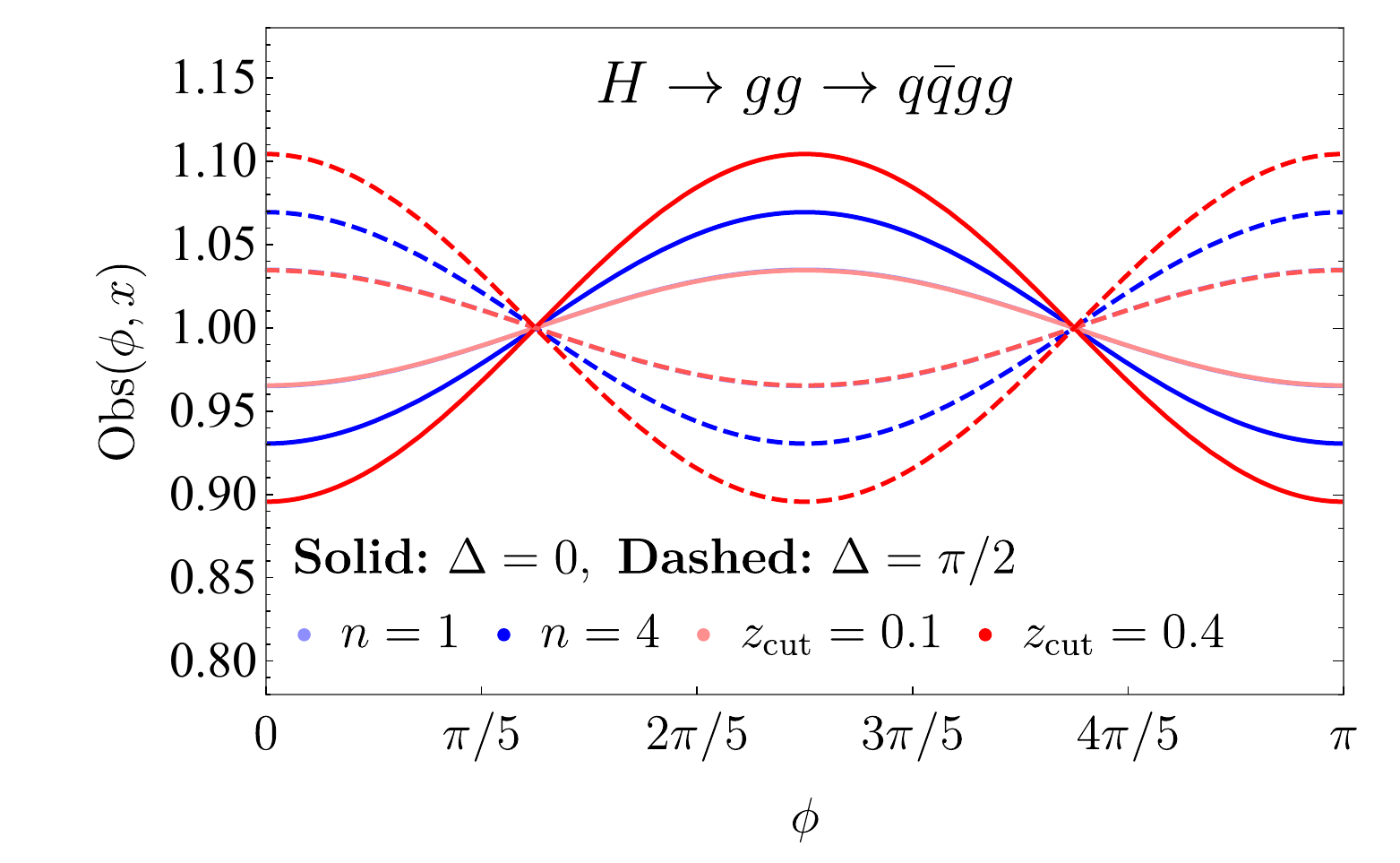} \\
        \includegraphics[width=0.46\textwidth]{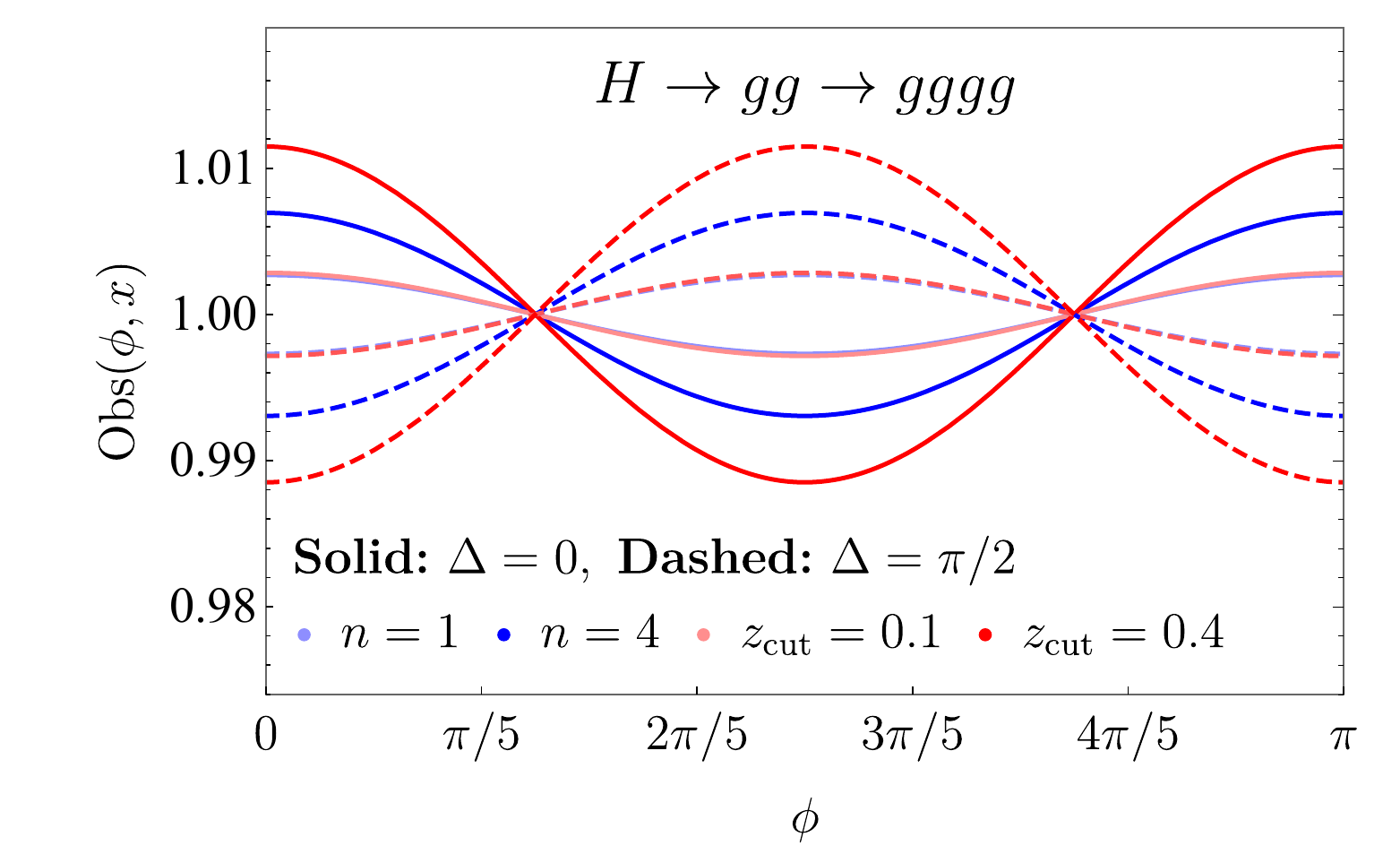} & \includegraphics[width=0.46\textwidth]{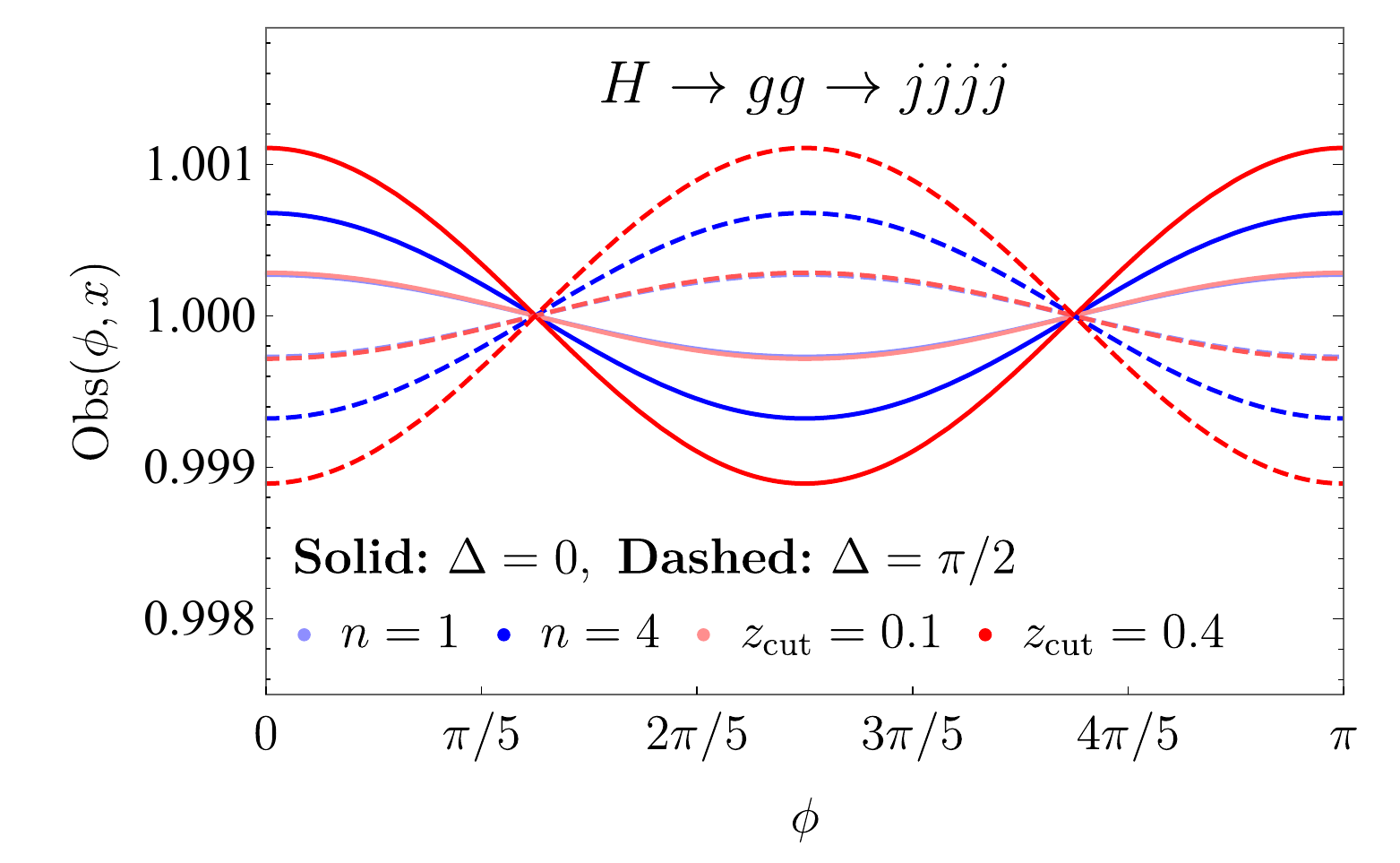}
    \end{tabular}
    \caption{${\rm Obs}(\phi, x)$ for the Lund observable with $z_{\rm cut}=0.1$ and $z_{\rm cut}=0.4$, as well as for the $\rm E4C$ observable with $n=1$ and $n=4$. Results for both $\Delta=0$ ($\mathcal{CP}$-even) and $\Delta=\pi/2$ ($\mathcal{CP}$-odd) in different splitting mode are presented. Since $A_{\text{E4C}}(n=1)$ and $A_{\text{Lund}}(z_{\text{cut}}=0.1)$ are very close, the corresponding curves overlap. 
    }
    \label{fig:calculation_z}
\end{figure*}

By convoluting the correlation strength of each splitting mode with its corresponding splitting probability, where approximately 93\% are $gggg$ modes, 7\% are $q\bar{q}gg$ modes, and 0.1\% are $qq\bar{q}\bar{q}$ modes, we obtain the correlation strength $A_{\text{Lund/E4C}}(x)$ for the inclusive channel $jjjj$, as listed in \tab{correlation}. The resulting $A_{\text{Lund/E4C}}(x)$ is significantly smaller than that of the individual splitting modes, due to partial cancelations between the two dominant contributing channels: $gggg$ and $q\bar{q}gg$, which have correlations of opposite sign. The distribution ${\rm Obs}(\phi, x)$ for the Lund observable and the $\rm E4C$ are further shown in \fig{calculation_z}, considering both cases $\Delta = 0$ and $\pi$, for different splitting modes. 

Among all splitting modes, the case $z_{\rm cut}=0.4$ shows a relatively large $A_{\text{Lund}}(x)$. However, this enhancement comes at the cost of reduced statistics due to the $z > z_{\rm cut}$ requirement. Table~\ref{tab:zcut_efficiency} highlights the efficiency $\epsilon_{z}$  of imposing this requirement for $z_{\rm cut}=0.4$ and $z_{\rm cut}=0.1$, which underscores the trade-off between maximizing correlation strength and retaining sufficient event rates.

\begin{table}[h!]
\centering
\begin{tabular}{|c|c|c|c|c|}
\hline
$\epsilon_{z}$ & $qqqq$  & $gggg$ & $q\bar{q}gg$ & $jjjj$ \\
\hline
$z_{\rm cut}=0.1$ & 0.64 & 0.32 & 0.45 & 0.33 \\
\hline
$z_{\rm cut}=0.4$ & 0.07 & 0.02 & 0.01 & 0.02 \\
\hline
\end{tabular}
\caption{Cut efficiency $\epsilon_{z}$ for $z_{\rm cut}=0.1$ and $z_{\rm cut}=0.4$ when considering Lund observable.}
\label{tab:zcut_efficiency}
\end{table}

\section{Phenomenology at future lepton colliders}
\label{sec:pheno}
We investigate the spin correlation between the two gluons from Higgs decay. For a future lepton collider operating at a center-of-mass energy of $\sqrt{s} = \unit[250]{GeV}$, we consider the dominant Higgs production channel $e^+ e^- \rightarrow Z H$ with a total cross section of $\sigma_{ZH} = 204.7 \ \text{fb}^{-1}$, and the leptonic decay channels of $Z\to l\bar{l}$ with $l\bar{l}$ including $e^+e^-$, $\mu^+\mu^-$ and the invisible $\nu\bar{\nu}$ channels. For these decay modes, the hadronic activities are purely from the Higgs decay. We neglect non-Higgs backgrounds such as $ZZ$, assuming they can be efficiently suppressed by reconstructing the invariant mass of the Higgs boson from the hadronic activities. In the $Z\to \nu\bar{\nu}$ channel, $H\to gg$ can be separated from $H\to b\bar{b}$ and $H\to VV^\ast\to jjjj$ with an efficiency of $\epsilon_{gg} = 90\%$, and a mistag rate of approximately 3\%, using event-level analysis techniques~\cite{Li:2020vav}. As a result, backgrounds from such decay modes of Higgs boson are subdominant and will be neglected after applying the cutting efficiency $\epsilon_{gg}$. A similar level of discrimination can be extended to the charged lepton decay channels of $Z$. The kinematics of Higgs boson can be reconstructed from the gluon jets which allows boosting to the Higgs rest frame for the construction of Lund observables and $ {\rm E4C}$.

The number of signal event after the above selection for each splitting mode is given by:
\begin{align}
    N_{X} & =  \sigma_{ZH} \mathcal{L} \, {\rm Br}({Z\to {l\bar{l}}}) \, {\rm Br}(H\rightarrow gg) P_{gg\rightarrow X}\, \epsilon_z\,\epsilon_{\rm cut} \epsilon_{gg},
\end{align}
with ${\rm Br}(Z\to l\bar{l})\simeq 0.27$, ${\rm Br}(H \to gg) = 0.0857$ and $P_{gg\rightarrow X}$ corresponding to the probability of $gg$ splitting to $X$. $\epsilon_z$ is only applied for the Lund observable with $z_{\rm cut}$ applied. We use $\epsilon_{\rm cut}$ to account for the minimal kinematic requirements for a jet (lepton) to be reconstructed at the detector, which are realized by requiring the corresponding parton (lepoton) to have $p_T \geq \unit[20(10)]{GeV}$, $\eta < 5(2.5)$ and $\Delta R > 0.4$ between these particles. Such requirements implemented with MadGraph lead to $\epsilon_{\rm cut} \simeq 0.63$. 

\subsection{Spin correlation measurement}
With $\Delta = 0$, we first examine the hypothesis that the spins are uncorrelated. Under this assumption, ${\rm Obs}(\phi, x)$ is expected to be uniformly distributed in $\phi$. Using $M = 5$ bins uniformly distributed in $\phi$ from 0 to $\pi$ and assuming each splitting modes at jet level can be fully separated, the significance for excluding the uncorrelated hypothesis is given by:
\begin{align}
    q  = \sqrt{-2\sum_{i=1}^{M} N_i^{\rm obs}-N_i+N^{\rm obs}_i\log \frac{N_i}{N^{\rm obs}_i}},
    \label{eq:significance}
\end{align}
where $N^{\rm obs}_i$ corresponds to the SM prediction with spin correlations and $N_i$ is signal event number predicted by the hypothesis to be tested at $i$-th bin. For the un-correlated hypothesis, we have $N_i=N_{\rm tot}/5$ where $N_{\rm tot}$ is the total event number integrating over all bins. The significance obtained at various choices of $x$ for two benchmark values of the luminosity $\mathcal{L}=5.6\ \text{ab} ^{-1}$ and $\mathcal{L}=20\ \text{ab}^{-1}$ \cite{CEPC:2022uwl} is shown in \tab{exclude_flat_E} and \tab{exclude_flat_E-20}, respectively.
\begin{table}[h!]
\centering
\begin{tabular}{|c|c|c|c|c|}
\hline
$q$ & $q\bar{q}q\bar{q}$ & $gggg$ & $q\bar{q}gg$ & $jjjj$\\
\hline
$z_{\rm cut}=0.1$ & 1.3608 & 0.1661 & 0.6688 & 0.0175 \\
\hline
$z_{\rm cut}=0.4$ & 0.6847 & 0.1047 & 0.3646 & 0.0108  \\
\hline
$n=1$ & 1.836 & 0.2795 & 1.007 & 0.0292  \\
\hline
$n=4$ & 2.8812 & 0.7204 & 2.005 & 0.0731  \\
\hline
\end{tabular}
\caption{Significance to exclude uncorrelated hypothesis using the choices of ${\rm E4C}(\phi, n=1)$ and ${\rm E4C}(\phi, n=4)$, and the Lund observable with $z_{\rm cut}=0.1$ and $z_{\rm cut}=0.4$, at $\mathcal{L}=5.6\ \text{ab}^{-1}$.}
\label{tab:exclude_flat_E}
\end{table}

\begin{table}[h!]
\centering
\begin{tabular}{|c|c|c|c|c|}
\hline
$q$ & $q\bar{q}q\bar{q}$ & $gggg$ & $q\bar{q}gg$ & $jjjj$ \\
\hline
$z_{\rm cut}=0.1$ & 2.517 & 0.3138 & 1.2639 & 0.0331 \\
\hline
$z_{\rm cut}=0.4$ & 1.2939 & 0.1979 & 0.6891 & 0.0248  \\
\hline
$n=1$ & 3.4697 & 0.5283 & 1.9031 & 0.0551  \\
\hline
$n=4$ & 5.4449 & 1.3614 & 3.7891 & 0.1381 \\
\hline
\end{tabular}
\caption{Same as \tab{exclude_flat_E}, with the luminosity changed to $\mathcal{L}=20\ \text{ab}^{-1}$.}
\label{tab:exclude_flat_E-20}
\end{table}

We observe that the significance of excluding the uncorrelated hypothesis is generally higher when using E4C compared to the Lund observable, due to the reduced statistics from applying $z_{\rm cut}$. Furthermore, in the limit of perfect separation between distinct splitting modes at jet level, future lepton colliders can exclude the uncorrelated hypothesis in either $q\bar{q}q\bar{q}$ or $q\bar{q}gg$ splitting mode at 95\% C.L.. The inclusive mode $jjjj$, while offering the largest statistical sample, shows negligible power to distinguish the correlated from the uncorrelated hypothesis, given its tiny correlation strength from the cancellations between different modes. 

As a ballpark estimate of the potential improvement from sub-jet tagging in evading the cancellation of the correlation strength in the inclusive \(jjjj\) mode, we consider applying a cut on the classification score \({\rm score}(g\to q\bar{q})\) introduced in~\cite{CMS-PAS-SMP-25-006}, which was designed to distinguish the splitting channels \(g\to gg\) and \(g\to q\bar{q}\). Using the efficiencies reported in Fig. 1 of Ref. \cite{CMS-PAS-SMP-25-006}, we estimate the modified significance by rescaling the inclusive \(jjjj\) sensitivity according to the surviving event fraction and the enhanced purity of the \(q\bar{q}gg\) component after the cut. We find that applying a cut \({\rm score}(g\to q\bar{q})>0.35\), which increases the purity of the \(q\bar{q}gg\) component from \(0.07\) to \(0.24\) and thus is assumed to significantly alleviates the cancellation effect, improves the estimated significance from \(0.1381\) to \(1.3422\), corresponding to roughly an order-of-magnitude enhancement in sensitivity. Since the tagging strategy in~\cite{CMS-PAS-SMP-25-006} targets secondary splittings, while our setup focuses on subjets originating from the first splitting, further improvements may be achievable with dedicated subjet-tagging techniques optimized for our analysis.
\begin{figure*}
    \centering
    \begin{tabular}{cc}
        \includegraphics[width=0.46\textwidth]{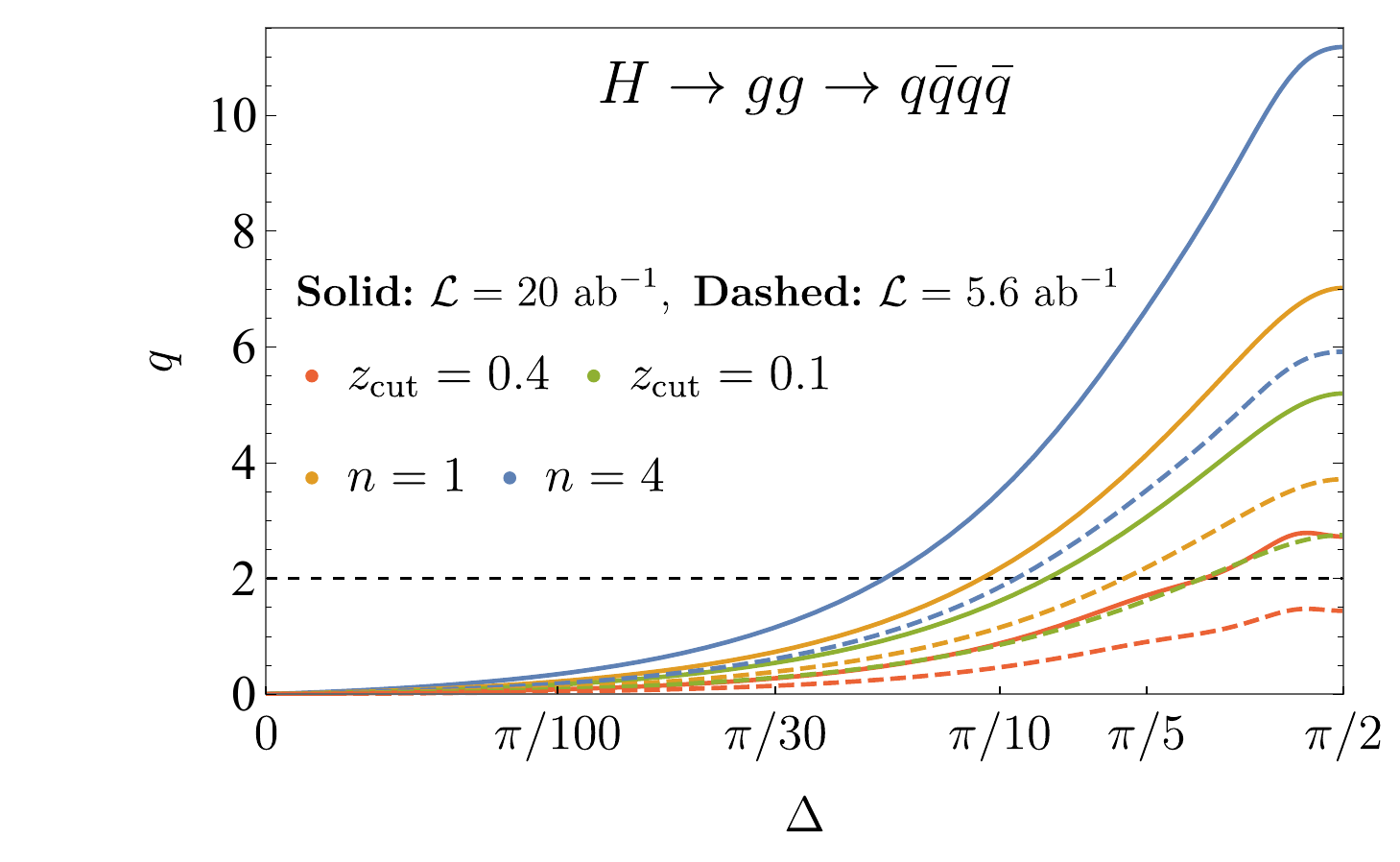} & \includegraphics[width=0.46\textwidth]{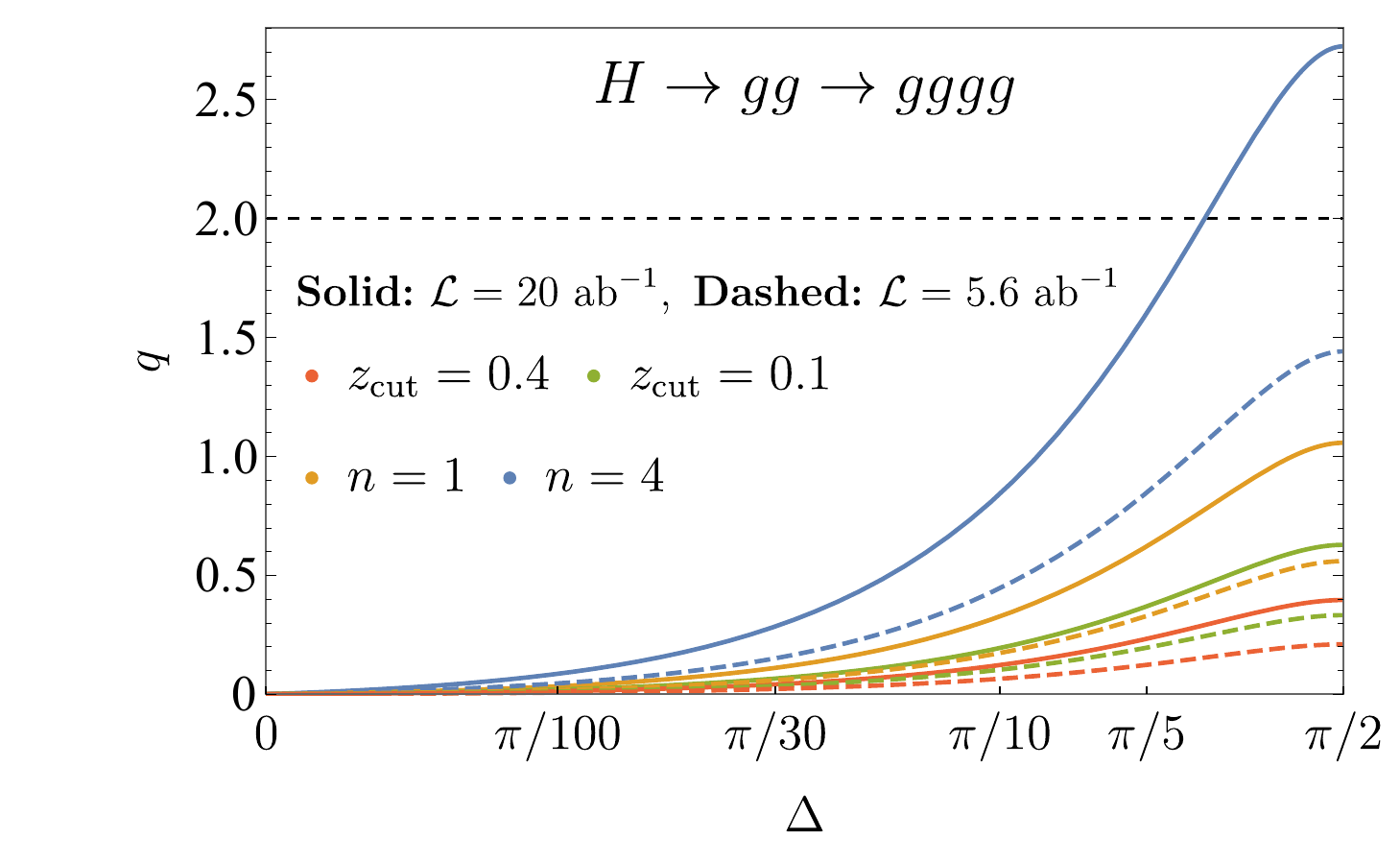} \\
        \includegraphics[width=0.46\textwidth]{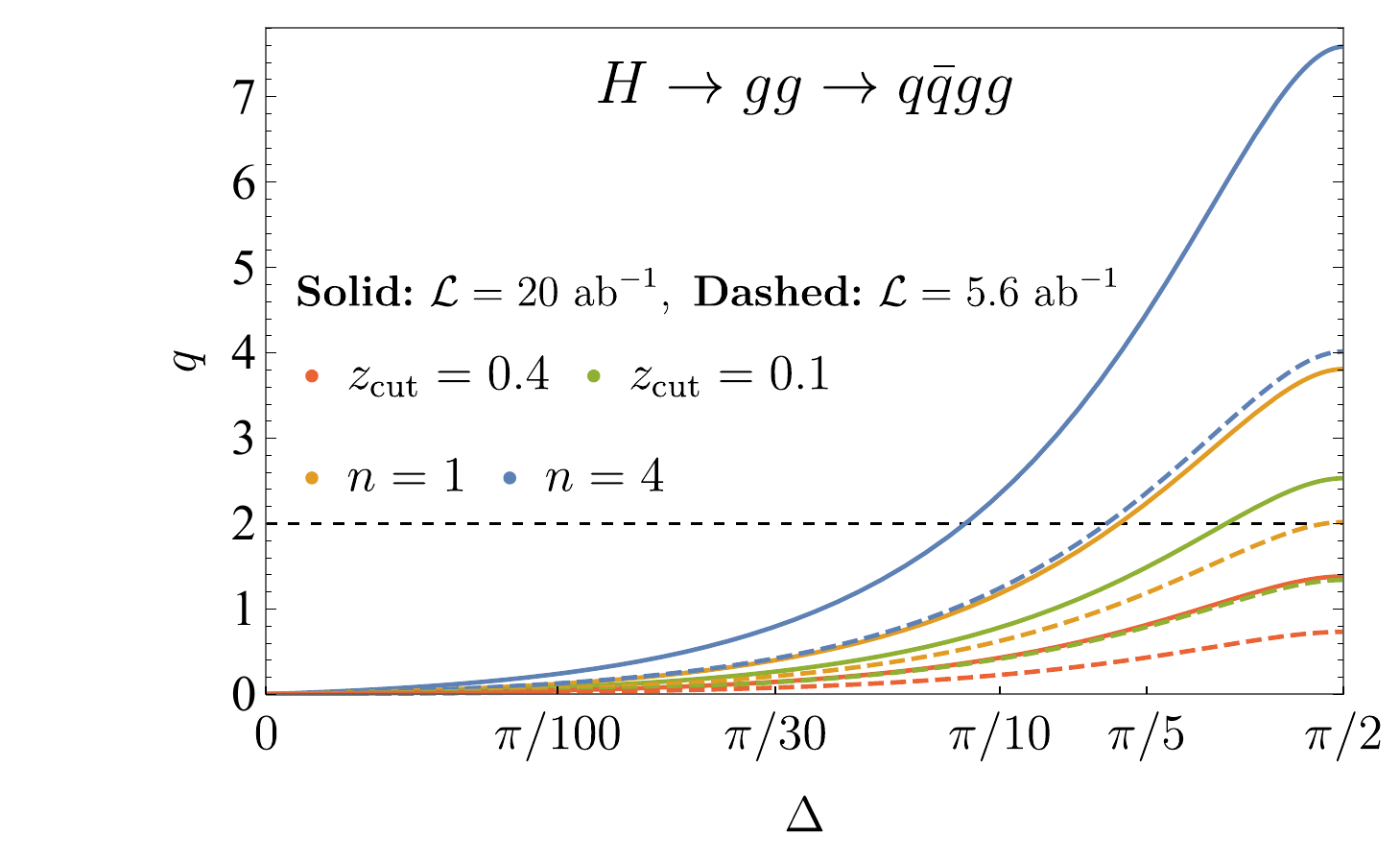} & \includegraphics[width=0.46\textwidth]{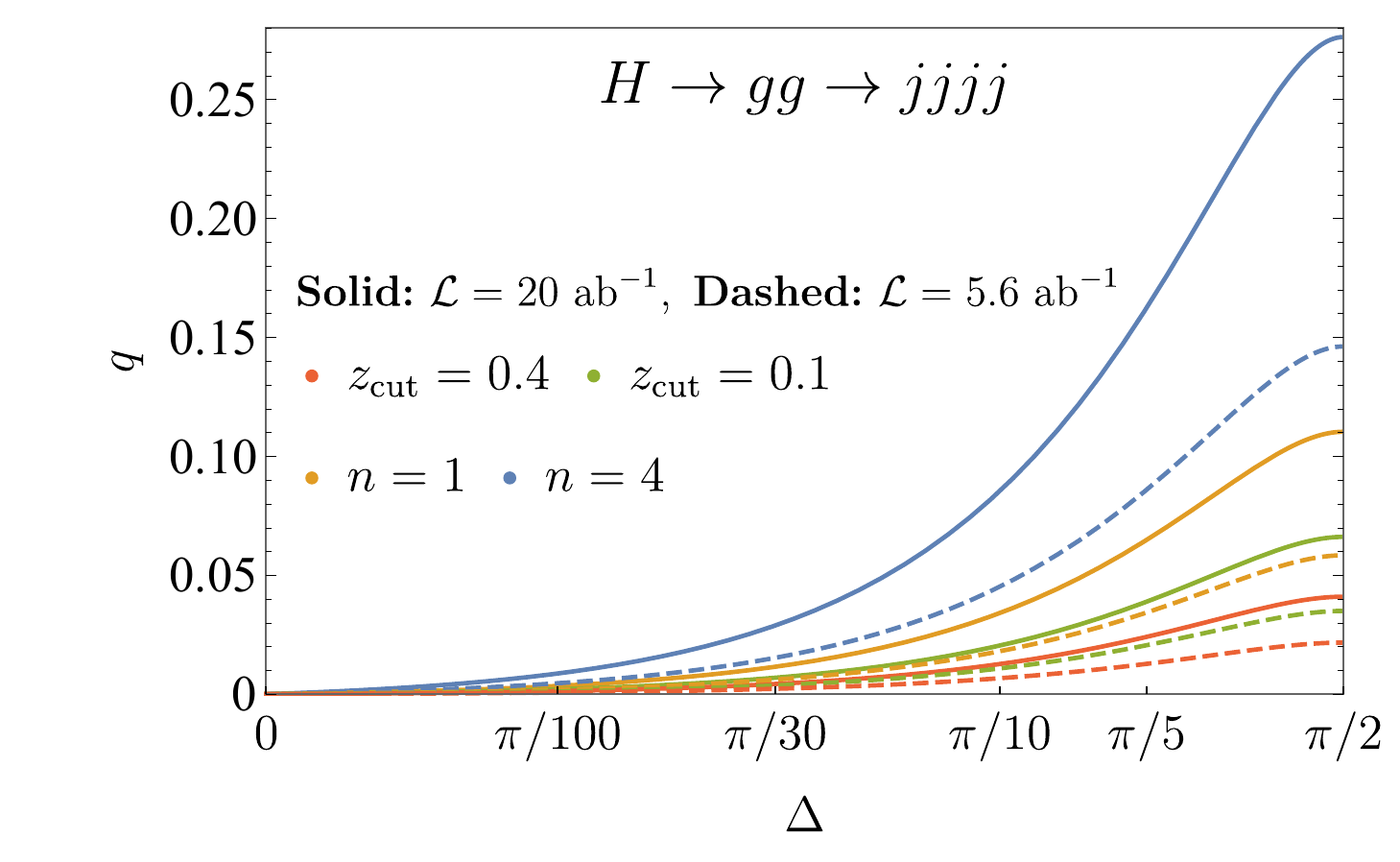}
    \end{tabular}
    \caption{The significance $q$ to exclude a $\mathcal{CP}$ mixing hypothesis with mixing angle $\Delta$, at $\mathcal{L}=20\ \text{ab}^{-1}$ and $5.6\ \text{ab}^{-1}$.}
    \label{fig:significance}
\end{figure*} 
\subsection{$\mathcal{CP}$ measurement}
As Lund observable and E4C are $\mathcal{CP}$-sensitive observables, we further explore their potential in probing the $\mathcal{CP}$ properties of the \(Hgg\) coupling at future lepton colliders. While indirect constraints on the $\mathcal{CP}$-mixing angle in the $Hgg$ coupling can be derived from neutron electric dipole moment (nEDM) measurements~\cite{Kley:2021yhn,Haisch:2019xyi,Cirigliano:2019vfc}, colliders enable direct probes of the $\mathcal{CP}$ property of $Hgg$ coupling. At the LHC, such probes rely on the azimuthal angle difference between the two associated jets in the \(H+2j\) production channel~\cite{Bahl:2023qwk,Englert:2019xhk,Bernlochner:2018opw}. In contrast, the Lund observable/E4C constructed from Higgs decay products provides a complementary and potentially first direct probe of the $\mathcal{CP}$ property of $Hgg$ coupling at lepton colliders.
For a given nonzero $\mathcal{CP}$-mixing angle \(\Delta\), we evaluate the exclusion sensitivity by computing the statistical significance using \eq{significance}, where \(N_i\) denotes the expected number of events in the \(i\)-th bin for the corresponding \(\Delta\) hypothesis. The sensitivities are shown in \fig{significance} for all four splitting modes, where the horizontal dashed line indicates the exclusion limit with $q =2$. We observe that for the best scenario of using ${\rm E4C}(\phi, n=4)$ in the $q\bar{q}q\bar{q}$ channel, a $\mathcal{CP}$-mixing angle as small as $\Delta\approx 0.03\pi$ can be excluded at future lepton colliders with $\sqrt{s} = 240$GeV and $\mathcal{L}=20\ \text{ab}^{-1}$ data. In comparison, the $\mathcal{CP}$-sensitive observable in the \(H+2j\) production channel has not yet demonstrated sensitivity to the $\mathcal{CP}$-mixing angle using the current dataset with \(\mathcal{L} = 139~\mathrm{fb}^{-1}\) LHC~\cite{Bahl:2023qwk}. Nevertheless, the $\mathcal{CP}$-mixing angle can be constrained to \(\Delta \lesssim 0.23\pi\) through a global fit including both \(H G^a_{\mu\nu}G^{a,\mu\nu}\) and \(H G^a_{\mu\nu}\tilde{G}^{a,\mu\nu}\) operators~\cite{Bahl:2023qwk}. For HL-LHC with \(\mathcal{L} = 3000~\mathrm{fb}^{-1}\), the same global fit is expected to improve the constraint to about \(\Delta \lesssim 0.11\pi\)~\cite{Bahl:2023qwk}.

\section{Conclusions}
\label{sec:conclusion}
In this work, we studied observables sensitive to gluon spin correlations in Higgs decays by looking at jet substructure observables. Two classes of observables were investigated: the four-point energy correlator (E4C) with different energy-weighting power and a Lund-plane-based variable. While the Lund observable requires a cut on subjet energy fraction to ensure IR safety, as well as to enhance the sensitivity to gluon spin correlations, it simultaneously reduces the available event statistics. Taking these competing effects into account, we find that the E4C observable with an energy-weighting power of \( n = 4 \) provides the highest sensitivity to gluon spin correlations at the parton level. 

Assuming ideal discrimination between different gluon splitting modes at the jet level, gluon spin correlations can be observed at future lepton colliders operating at \( \sqrt{s} = 240~\mathrm{GeV} \) with an integrated luminosity of \( \mathcal{L} = 5.6~\mathrm{ab}^{-1} \), particularly through the splitting channel \( gg \to q\bar{q}q\bar{q} \). Although the inclusive gluon splitting mode provides the largest event yield, its sensitivity is strongly reduced due to cancellations between contributions from \( g \to q\bar{q} \) and \( g \to gg \) splittings, resulting in a much weaker net correlation. Extending this framework to direct measurements of the $\mathcal{CP}$ structure of the \( Hgg \) coupling, we find that the \( gg \to q\bar{q}q\bar{q} \) channel can probe the $\mathcal{CP}$-mixing angle to \( \Delta \phi \simeq 0.03\pi \) at \( \sqrt{s} = 240~\mathrm{GeV} \) with an integrated luminosity of \( \mathcal{L} = 20~\mathrm{ab}^{-1} \).

The strategy of using energy-energy correlation can also be extended to potentially improve the intra-jet spin correlation measurement at LHC \cite{CMS-PAS-SMP-25-006}. Yet, this strategy extending to probe the \( b\bar{b} \) spin correlations and the $\mathcal{CP}$ structure of the \( Hb\bar{b} \) coupling in $H\to b\bar{b}$ decay is strongly limited, as the quark splittings into vector bosons do not preserve the initial quark spin correlations, in agreement with expectations from soft theorems. 

Future work will focus on the development of dedicated subjet-tagging techniques to enable separations of different splitting modes at jet-level. This will allow a full exploitation of the potential of the E4C observable at future lepton colliders for probing partonic spin correlations and for searching for new physics beyond the Standard Model.
\section*{Acknowledgement}

We would like to thank Shuo Han, Manqi Ruan, Bin Yan and Hongtao Yang for useful discussions. The authors gratefully acknowledge the valuable discussions and insights provided by the members of the Collaboration on Precision Tests and New Physics (CPTNP). This work is supported by the National Key
R\&D Program of China (Grant Nos. 2023YFA1609300, 2025YFA1614200), and also supported by the National Science Foundation
of China under Grant Nos. 12522509, 12305107, 12275173,
12425505. 

\bibliography{refs}

\end{document}